\documentclass[aps, prl, superscriptaddress, reprint, twocolumn,]{revtex4-2}
\usepackage{amsmath}
\usepackage{amsfonts}
\usepackage{amssymb}
\usepackage{gensymb}
\usepackage{graphicx}
\usepackage[utf8]{inputenc}
\usepackage[colorlinks=true, linkcolor=black, urlcolor=blue, citecolor=black, anchorcolor=black]{hyperref}
\usepackage{enumitem}
\setlist[itemize]{leftmargin=*}

\setlist{nosep}

\graphicspath{{"figures/"}}

\begin{document}
\title{Noisy detectors with unmatched detection efficiency thwart identification of single photon emitters via photon coincidence correlation}
\author{Darien J. Morrow}
\email{dmorrow@anl.gov}
\affiliation{Center for Nanoscale Materials, Argonne National Laboratory, Lemont, Illinois 60439, United States}

\author{Xuedan Ma}
\email{xuedan.ma@anl.gov}
\affiliation{Center for Nanoscale Materials, Argonne National Laboratory, Lemont, Illinois 60439, United States}
\affiliation
{Consortium for Advanced Science and Engineering, University of Chicago, Chicago, Illinois 60637, United States}
\affiliation
{Northwestern-Argonne Institute of Science and Engineering, 2205 Tech Drive, Evanston, IL 60208, USA}


\begin{abstract}
	Single photon emitters (SPEs) are often identified with a Hanbury Brown and Twiss intensity interferometer (HBTII) consisting of a 50:50 beamsplitter and two time-correlated single photon counters. For an SPE, the cross-correlation of photon arrival time between the two detectors, $g^{(2)}(\tau)$, shows a hallmark dip to zero at $\tau=0$. One common heuristic for identifying SPEs is by measuring a system to have $g^{(2)}(0)<1/2$. Here in, we use stochastic methods to simulate the case of a SPE observed with non-ideal detectors and optics. We show that identification of SPEs is thwarted when the detectors have asymmetric detection efficiency, and when detector background noise is more than half the true SPE signal rate. 
\end{abstract}

\maketitle

Single photon emitters (SPEs) play a crucial role in many extant and proposed quantum information technologies.\cite{Aharonovich_Toth_2016, OBrien_Vuckovic_2009}
Categorization of an emitter as a SPE is usually accomplished by measuring the degree of second-order temporal coherence, $g^{(2)}(\tau)$, of the emission. 
For photons described by Fock states, $|n\rangle$, of number $n$, the second-order coherence is\cite{Loudon_2000, Paul_1982}
\begin{align}
	g^{(2)}(\tau) &= \frac{\left\langle n^2\right\rangle - \left\langle n\right\rangle }{\left\langle n\right\rangle^2} \\
	&= 1 - \frac{1}{n} \text{ for } n\geq 1. \label{eq:fock}
\end{align}
If only one photon exists at a time then $g^{(2)}(\tau) = 0$, and if two photons exist at a time then $g^{(2)}(\tau) = 1/2$; this analysis implies that if an emitter is found to have $g^{(2)}(0) < 1/2$, that emitter must be a SPE.\cite{Michler_Imamoglu_2000, Hgele_Imamoglu_2008} 

\begin{figure}[!htbp]
	\centering
	\includegraphics[trim=0cm 1cm 1cm 1cm, clip,width=\linewidth]{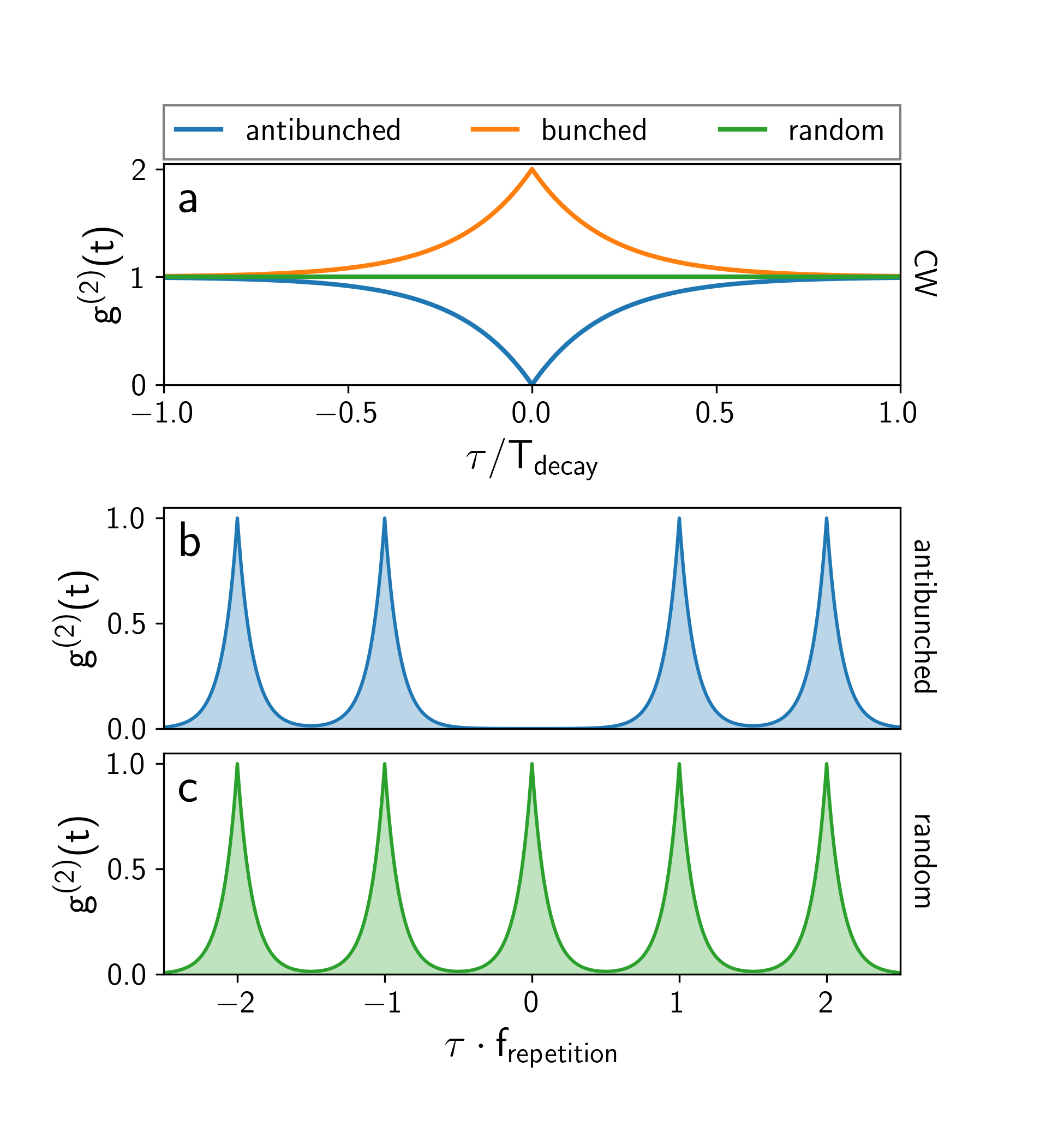}
	\caption{Simulation of photon antibunching data. (a) Cross-correlation from a HBTII where the length of one arm is swept while measuring fluorescence from a chromophore excited with a continuous wave (CW) source. (b,c) Cross-correlation of photon arrival times at two time-tagging detectors when measuring fluorescence from a chromophore excited with a pulsed source. (b) shows prominent antibunching in photon arrival time while (c) exhibits random (Poisson statistics) arrival times. 
	}\label{fig:g2} 
\end{figure}

In order to measure $g^{(2)}(\tau)$, most labs employ variants of the Hanbury Brown and Twiss intensity interferometer (HBTII) consisting of a 50:50 beamsplitter which feeds two detector arms equipped with time correlated single photon counters.\cite{Brown_Twiss_1956}
In terms of the long-time average intensity, $\bar{I}$, the normalized second order coherence is\cite{Loudon_2000, Paul_1982, Kimble_Mandel_1977} 
\begin{align}
	g^{(2)}(\tau) = \frac{\left\langle \bar{I}(t) \bar{I}(t+\tau)\right\rangle}{\bar{I}^2}, \label{eq:gint}
\end{align}
which has properties of $g^{(2)}(0) \geq 1$ for thermal emission and $g^{(2)}(0) = 1$ for coherent (random, Poissonian) emission.
Classical (not quantized) light exhibits bunching $g^{(2)}(0) \geq 1$ while quantum light exhibits antibunching $g^{(2)}(0) < 1$.
Experimentally, a conspicuous dip of $g^{(2)}(\tau)$ near $\tau=0$ heralds the existence of a quantized phenomena.\cite{Walls_1979}
\autoref{fig:g2} shows simulated correlation functions when the emitters are excited by continuous (\autoref{fig:g2}a) or pulsed (\autoref{fig:g2}b,c) sources.

Traditionally, background counts originating from scattered excitation light, or detector shot noise are understood to uniformly increase the value of $g^{(2)}(\tau=0)$ measured by a HBTII.\cite{Kimble_Mandel_1977, Brouri_Grangier_2000, Kuhn_Rempe_2002, Kurtsiefer_Weinfurter_2000}
Simplistically, if a SPE measurement is saturated with truly random background noise, the coherent case of $g^{(2)}(0) = 1$ results.\cite{Brouri_Grangier_2000}  
While ubiquitous, background noise is not the only nonideality which can plague SPE measurements.
In our lab, we have identified three phenomena which can possibly complicate the measurement of SPEs with our HBTIIs:
\begin{itemize}
	\item high levels of random background noise logged by both photon counting modules,
	\item photon counting modules with different quantum efficiencies or beamsplitters which do not evenly feed the counting modules,
	\item and photon counting modules with intrinsically different background count rates. 
\end{itemize}
This work computationally demonstrates how these three phenomena can conspire to make SPE identification impossible. 

\begin{figure}[!htbp]
	\centering
	\includegraphics[trim=0cm 0cm 0cm 0cm, clip,width=.8\linewidth]{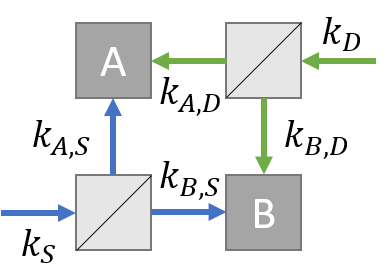}
	\caption{Idealized Hanbury Brown and Twiss intensity interferometer. A and B are detectors which detect fluorescence signal at a rate of $k_{A,S}$ and $k_{B,S}$, respectively. Dark (background) counts are recorded at a rate of $k_{A,D}$ and $k_{B,D}$, respectively.
	}\label{fig0} 
\end{figure}

To model the role of these three nonidealities, we imagine an impulsive, pulsed source exciting a single, perfect SPE which emits exactly one photon per excitation pulse resulting in a photon emission rate of $k_{S}$.
As diagrammed in \autoref{fig0}, the stream of photons from the SPE is split into two streams, $k_{A,S}^\prime$ and $k_{B,S}^\prime$, which excite two detectors, $A$ and $B$, respectively.
The beamsplitter is assumed to be lossless,
\begin{align}
	k_{S} = k_{A,S}^\prime + k_{B,S}^\prime.
\end{align}   
If the detectors are asymmetrically fed by the beamsplitter (e.g. it is not 50:50) or if they have different quantum efficiencies, $\Xi$, then the effective count rate is
\begin{align}
	k_{A,S} &= \Xi\left(\frac{1}{2}+\xi \right)k_{S},\\
	k_{B,S} &= \Xi\left(\frac{1}{2}-\xi \right)k_{S},
\end{align}
for $\xi \in (-1/2,1/2)$ and $\Xi \in (0,1]$. 
For this work, we let $\Xi=1$.

To add noise to our model, we define a total background count rate, $k_D$, in reference to the total signal rate,
\begin{align}
	k_D = \sigma \Xi k_S,
\end{align}
with $\sigma$ being the \emph{noise factor} which connects the two rates. 
Just like the signal rate, the background rate is losslessly split and fed into the two detectors 
\begin{align}
	k_{A,D} &= \left(\frac{1}{2}+\chi \right)k_{D},\\
	k_{B,D} &= \left(\frac{1}{2}-\chi \right)k_{D},
\end{align}
with $\chi \in [-1/2,1/2]$ being the background rate asymmetry between the two detectors. 

This system of rates encodes in scalar factors the three detrimental phenomena previously listed:
\begin{itemize}
	\item $\sigma$ describes the \emph{background count rate} relative to the signal count rate, it can be experimentally controlled by e.g. cooling the detectors;\cite{Hofbauer_Zimmermann_2018, Richardson_Henderson_2009, Kang_Risk_2003, Spinelli_Lacaita_1997}
	\item $\xi$ describes the \emph{asymmetry in signal rates} between the two detectors, it can be experimentally controlled by e.g. adding neutral density filters in front of one of the detectors;
	\item $\chi$ describes the \emph{asymmetry in background rates} between the two detectors, it is hard to experimentally control because it is caused by e.g. differences in trap concentration in the active elements of the detectors.\cite{Kang_Risk_2003, Spinelli_Lacaita_1997}
\end{itemize}

A HBTII instrument yields two streams of photons whose arrival times at the two detectors are recorded.
We write the digital data streams from the detectors A and B as $A(t)$ and $B(t)$ where $t$ is laboratory time.
$A(t)=1$ ($B(t)=1$) if a photon hits detector A (B) at time $t$, while $A(t)=0$ ($B(t)=0$) if no photon hits detector A (B) at time $t$.
The normalized cross-correlation between the two data streams,\footnote{For continuous functions the cross-correlation, $z$, is given by,$z(\tau) = (x \star y)(\tau) \equiv \int_{-\infty}^{\infty} \overline{x(t)} y(t+\tau)dt$ where $\overline{f(t)}$ is the complex conjugate of $f(t)$. For finite, 1D arrays the cross-correlation is instead	$z[k] = (x \star y)(k - N + 1) = \sum_{l=0}^{||x||-1}x_l \overline{y_{l-k+N-1}}$ for $k = 0, 1, \dots, ||x|| + ||y|| - 2$ in which $||x||$ is the length of $x$, $N = \max(||x||,||y||)$ and $y_m$ is zero-padded such that when $m$ is outside the range of $y$ then $y_m=0$.}
\begin{align}
	g^{(2)}(\tau) = \frac{ A(t) \star B(t)}{\left|\left|A(t) \star B(t)\right|\right|},
\end{align}
is analogous to \autoref{eq:gint}.
For experiments conducted with pulsed excitation sources it is common to quantify $g^{(2)}(0)$ using integrals over specific pulse periods ($T$) of the excitation laser 
\begin{align}
	g^{(2)}(0) \approx \frac{2n
		\int_{-T/2}^{T/2}(A \star B)\textrm{d}t}{
		\int_{-nT-T/2}^{-T/2}(A \star B)\textrm{d}t + \int_{T/2}^{nT+T/2}(A \star B)\textrm{d}t} \label{eq:g0int}
\end{align}
where $n$ is a small, positive integer.
This is the definition of $g^{(2)}(0)$ which we will use for the remainder of this work.

In order to simulate measurement of a SPE, we assume impulsive excitation, emission, and detection of the SPE with unity quantum efficiency.
This Dirac-delta distribution nature of dynamic processes allows us to write the data streams, $A$ and $B$, as simple 1D numerical arrays whose index is identical to the laser pulse period. 
To populate $A$ and $B$, for each array index, $i$, a random number, $\rho$, is drawn from a uniform distribution, $\rho \in [0,1)$ and the arrays are written as:
\begin{itemize}
	\item If $\rho_i < 0.5 + \xi$ then $A[i]=1$ and $B[i]=0$. 
	\item If instead $\rho_i > 0.5 + \xi$ then $A[i]=0$ and $B[i]=1$
\end{itemize}
To add noise into the data stream, the arrays are updated using two more random numbers, $\rho^\prime \in [0,1)$ and $\rho^{\prime\prime}\in [0,1)$:
\begin{itemize}
	\item If $\rho_i^\prime < 0.5 + \chi$ then $A[i]\stackrel{+}{=}\sigma$ if not, $A[i]\stackrel{+}{=}0$, 
	\item If $\rho_i^{\prime\prime} > 0.5 + \chi$ then $B[i]\stackrel{+}{=}\sigma$ if not, $B[i]\stackrel{+}{=}0$,
\end{itemize}
with the generation of $\rho^\prime$ and $\rho^{\prime\prime}$ being uncorrelated to ensure the background noise is not correlated between the two detectors. 
This iterative array population is accomplished for arrays of size $N$.
$N$ is both the number of pulse periods over which the experiment has run as well as the number of single photon emission events.

\begin{figure}[!htbp]
	\centering
	\includegraphics[trim=0cm 1cm 1cm 1.5cm, clip,width=\linewidth]{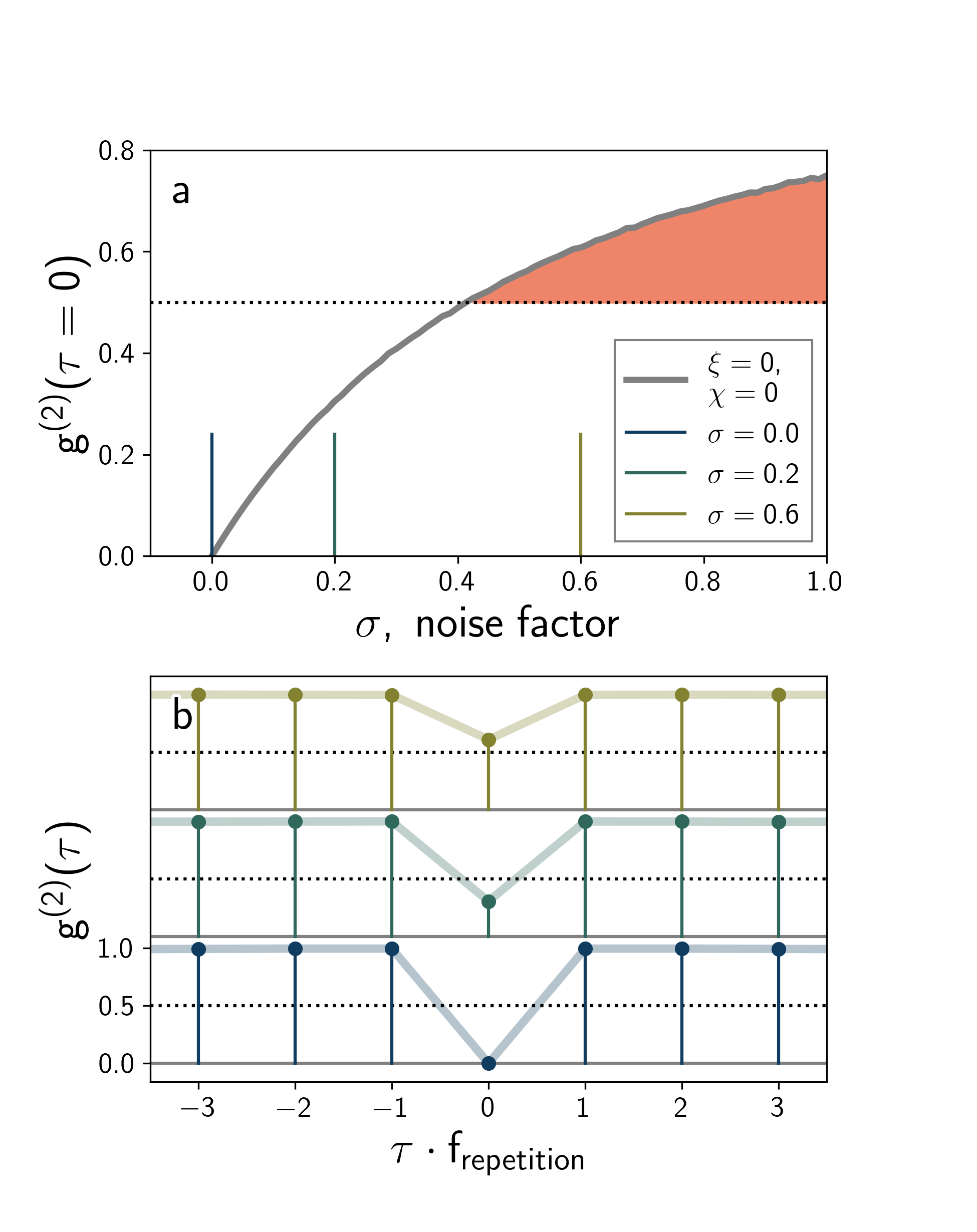}
	\caption{$g^{(2)}(\tau)$ as noise factor is increased for $\xi=\chi=0$ ($N=20000$). (a) shows that $g^{(2)}(\tau=0)$ increases as noise is increased, values of $g^{(2)}(0)>1/2$ are highlighted by red fill. (b) shows $g^{(2)}(\tau)$ for three different noise scales (from bottom to top: $\sigma=0$, $\sigma=0.2$, and $\sigma=0.6$) emphasizing the Dirac delta nature of the simulation in comparison to the finite distributions of \autoref{fig:g2}b,c.
	}\label{fig1} 
\end{figure}

We first simulate the role that an increase in background noise plays in determining the value of $g^{(2)}(\tau)$ when noise and signal counts are symmetrically distributed across the two detectors ($\xi=\chi=0$).
\autoref{fig1}a shows that with no background noise, $\sigma=0$, $g^{(2)}(0)=0$.
As noise increases, $g^{(2)}(0)$ smoothly increases until it passes  $g^{(2)}(0)=0.5$ at around $\sigma=0.4$; above this point the simulation indicates that detector noise makes it impossible to identify SPEs using only the metric of $g^{(2)}(0)<0.5$.
\autoref{fig1}b likewise shows how the temporal structure of $g^{(2)}(\tau)$ smoothly evolves as background noise increases---note the resemblance of \autoref{fig1}b to a Dirac comb distribution.
\autoref{fig1} demonstrates that our model recovers the common observation that background noise pushes up the minimum achievable value of $g^{(2)}(0)$.\cite{Kimble_Mandel_1977, Brouri_Grangier_2000, Kuhn_Rempe_2002, Kurtsiefer_Weinfurter_2000}  

\begin{figure}[!htbp]
	\centering
	\includegraphics[trim=.5cm 1cm .5cm 1.7cm, clip,width=\linewidth]{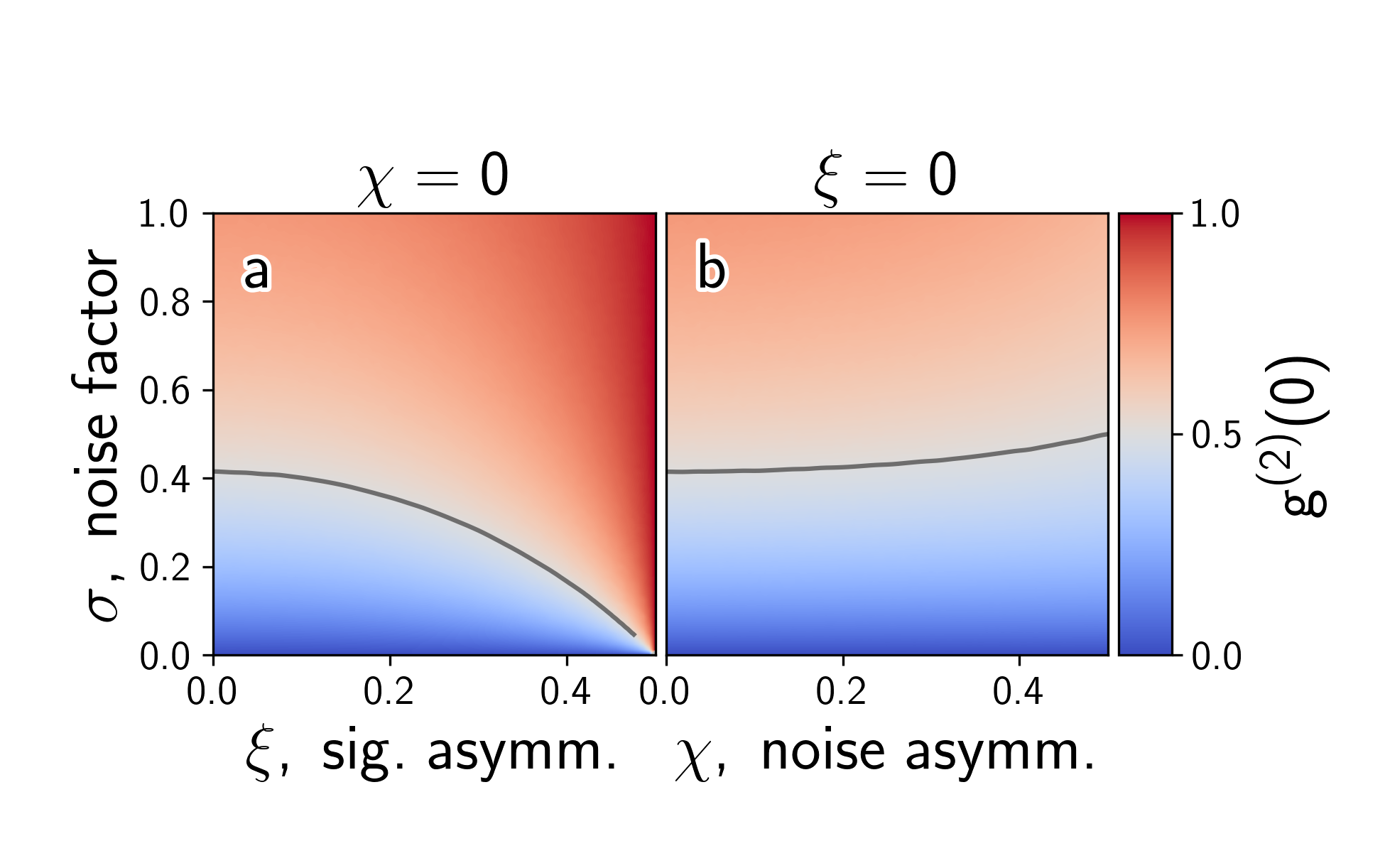}
	\caption{$g^{(2)}(0)$ as a function of background noise and detector asymmetry in signal (a) and noise (b). Note that the colormap is blue for values of $g^{(2)}(0)<0.5$ which traditionally is the metric of a single photon emitter. We draw a gray contour line at $g^{(2)}(0)=0.5$.
	}\label{fig2} 
\end{figure}

Next we model how asymmetry in the detection efficiency or noise distribution affects the observable values of $g^{(2)}(0)$. 
\autoref{fig2}a demonstrates that both background noise and detection efficiency asymmetry increase the observable value of $g^{(2)}(0)$.
As an example, for $\xi= 0.3$ (i.e. an 80:20 beamsplitter) a noise factor of a little more than $\sigma=0.2$ causes our ideal SPE to not be identified as a SPE using the $g^{(2)}(0)<0.5$ rule-of-thumb.

When there is no detection asymmetry, $\xi= 0$, increasing the asymmetry of noise, $\chi$, between the two detectors does not greatly affect the cutoff value of noise for $g^{(2)}(0)<0.5$ (\autoref{fig2}b).  
If only one detector has background noise ($\chi=0.5$) then the HBTII can still identify a SPE even with the noisy detector having the same signal and noise count rates. 
This effect may be rationalized by considering how a noise rate of $\sigma k_S$ on only one of the two detectors factors into \autoref{eq:g0int}, 
\begin{align}
	g^{(2)}(0) = \frac{0 + \sigma k_S }{k_S+\sigma k_S}.
\end{align}
When noise and signal rates at one detector are similar ($\sigma=0.5$, $k_{A,D}=k_{A,S}=k_{B,S}$, $k_{D,S}=0$) then $g^{(2)}(0) = 0.5$.
This observations yields a heuristic that SPE identification is severely hampered if the combined background noise rates of both detectors is greater than 50\% of the combined signal rates.

\begin{figure}[!htbp]
	\centering
	\includegraphics[trim=0cm .75cm 0cm 1.5cm, clip,width=\linewidth]{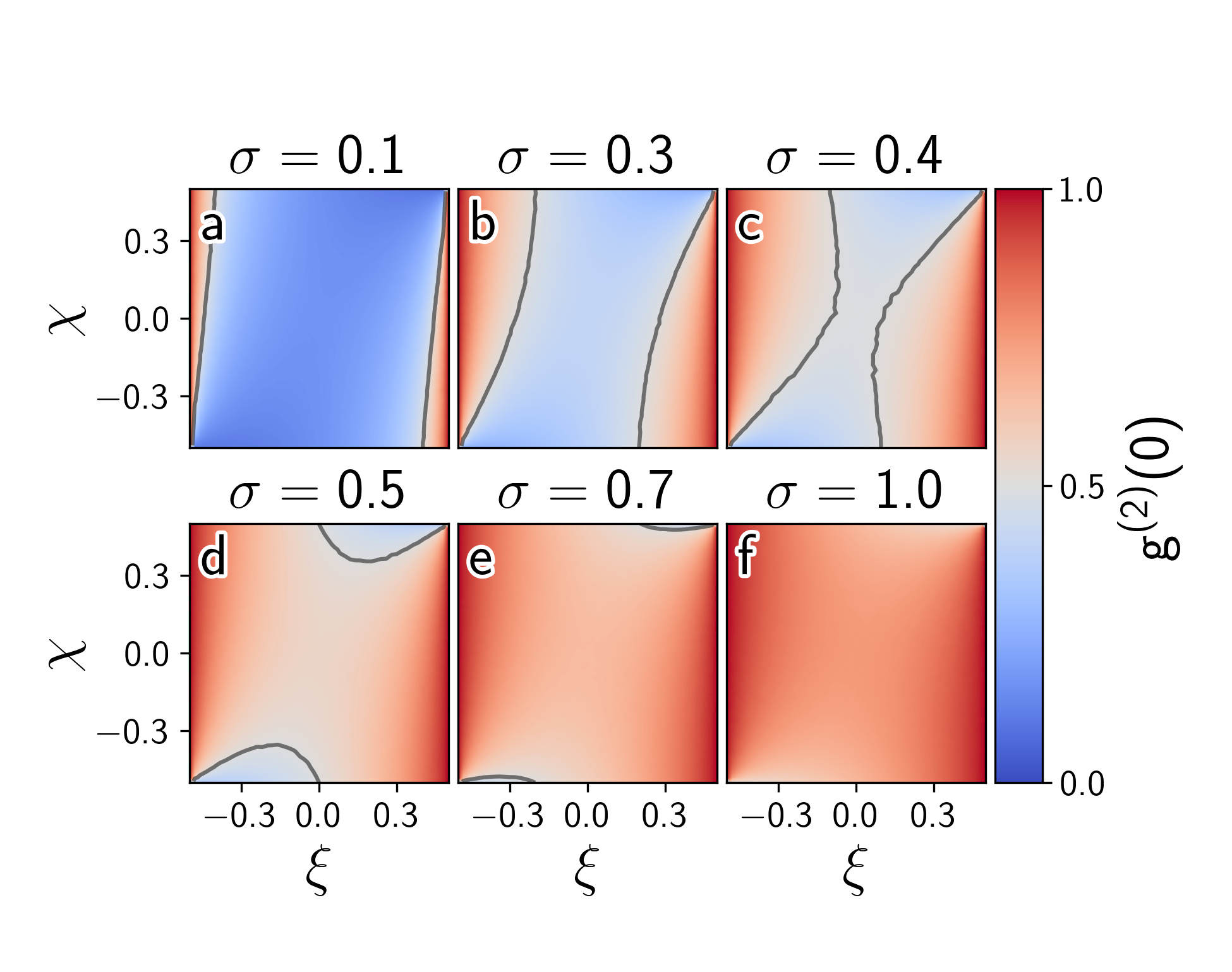}
	\caption{$g^{(2)}(0)$ as a function of asymmetry of noise and signal for the two detectors. The noise factor, $\sigma$, is varied from 0.1 to 1 in these figures.
	}\label{fig3} 
\end{figure}

In order to further investigate the inability to measure $g^{(2)}(0)<0.5$ when $\sigma\gtrsim 0.5$, we simulate how $g^{(2)}(0)$ changes jointly as a function of $\xi$ and $\chi$ while stepping through values of $\sigma$ (\autoref{fig3}).
\autoref{fig3}f shows that when $\sigma=1$ there are no values of $\xi$ and $\chi$ which yields $g^{(2)}(0)<0.5$.
As the noise factor decreases a region around the line $\xi=\chi$ begins to have $g^{(2)}(0)<0.5$.
At low background noise levels, all but the largest signal detection asymmetries yield $g^{(2)}(0)<0.5$. 

Taken together, this work allows us to bound the region of noise rates and detection asymmetry which can hamper SPE categorization with a HBTII.
Because this work assumes the platonic ideal of both SPEs and HBTIIs with exactly one photon emitted and detected per excitation pulse period, the derived heuristics should be taken as an upper-bound on the non-idealities which can exist and still allow for categorization of SPEs.  

For facile categorization of SPEs, the signal count rate at the individual detectors must be larger than the noise count rate for each detector.
Asymmetry in signal detection efficiency should be ardently avoided; larger amounts of background noise compounds the detrimental nature of asymmetric signal detection.
Conversely, asymmetry in background noise between the two detectors does not have a significant detrimental effect on a HBTII's ability to categorize SPEs.  
If one does have asymmetric signal and noise rates between the two detectors, it is best to match them so one detector has both the highest signal and background noise rates. 

Finally, we note that while $g^{(2)}(0)<0.5$ is a fundamental characteristic of a SPE (\autoref{eq:fock}), imperfections in the HBTII characterizing the SPE can make it impossible to categorize the SPE as such. 
Much like recent work which indicated that, within experimental limitations, $g^{(2)}(0)$ is a poor metric for categorizing the lasing threshold of microlasers,\cite{Carroll_Papoff_2021} here we stress that $g^{(2)}(0)<0.5$ is only a good identifier for SPEs when background noise is less than half the fundamental signal rate and detectors are matched in efficiency.

\hfill

This work was performed at the Center for Nanoscale Materials, a U.S. Department of Energy Office of Science User Facility, was supported by the U.S. DOE, Office of Basic Energy Sciences, under Contract No. DE-AC02-06CH11357.

\bibliography{database}
\end{document}